\documentclass[10pt,a4paper]{article}

\usepackage{times,cite}
\usepackage{graphicx}
\usepackage{epsfig,wrapfig}
\usepackage{epstopdf} 
\usepackage{xcolor}
\usepackage{fancyhdr}
\usepackage{amsmath, amssymb}

\usepackage[lmargin=2.5cm, rmargin=2.5cm,tmargin=3.50cm,bmargin=3.50cm]{geometry}
\usepackage{indentfirst}

\usepackage{titlesec}
\titleformat{\section}[hang]
  {\centering}{\thesection}{1ex}{\normalsize \textsc}
\titleformat{\subsection}[hang]
  {}{\thesubsection}{1ex}{\normalsize \textit}

\renewcommand{\thesection}{ \normalsize \textnormal{\Roman{section}.}}
\renewcommand{\thesubsection}{\normalsize \textnormal{\textsc{\textit{\Alph{subsection}.}}}}

\def\e{\begin{equation}}
\def\f{\end{equation}}
\def\_#1{{\bf #1}}

\def\.{\cdot}

\def\Re{{\rm Re\mit}}

\begin{document}

\title{\large \textbf{Semianalyitcal synthesis scheme for multifunctional metasurfaces on demand}}
%
\def\affil#1{\begin{itemize} \item[] #1 \end{itemize}}
\author{\normalsize \bfseries V. K. Killamsetty and \underline{A. Epstein}
\\ }
\date{}
\maketitle
\thispagestyle{fancy} 
\vspace{-6ex}
\affil{\begin{center}\normalsize Technion - Israel Institute of Technology,\\
 Andrew and Erna Viterbi Faculty of Electrical Engineering, Haifa 3200003, Israel\\
epsteina@ee.technion.ac.il
 \end{center}}

\begin{abstract}
\noindent \normalsize
\textbf{\textit{Abstract} \ \ -- \ \
We propose a comprehensive field-based semianalytical method for designing fabrication-ready multifunctional periodic metasurfaces (MSs). 
Harnessing recent work on multielement metagratings based on capacitively-loaded strips, we have extended our previous meta-atom design formulation to generate realistic substrate-supported printed-circuit-board layouts for anomalous refraction MSs. 
Subsequently, we apply a greedy algorithm for iteratively optimizing individual scatterers across the entire macroperiod to achieve multiple design goals for corresponding multiple incidence angles with a single MS structure.  
As verified with commercial solvers, the proposed semianalytical scheme, properly accounting for near-field coupling between the various scatterers, can reliably produce highly efficient multifunctional MSs on demand, without requiring time-consuming full-wave optimization. 
}
\end{abstract}

\section{Introduction}

Metasurfaces (MSs) are low-profile composites, consisting of closely-packed subwavelength polarizable elements (meta-atoms), arranged in regular dense lattices. Such structures have been attracting significant attention in recent years for their capability to implement complex functionalities across the electromagnetic spectrum \cite{glybovski2016metasurfaces}.
MSs are typically designed in a two-step process: first, the required homogenized constituent distribution is extracted from the desired field transformation (macro); next, the resolved abstract surface properties are discretized, and physical geometries are assigned to meta-atoms, realizing the prescribed local response at each point (micro) \cite{epstein2016huygens}.
At microwave and mm wave frequencies, meta-atoms are often implemented as cascaded impedance sheets [compatible with multilayer printed-circuit-board (PCB) configurations], and the microscopic design is carried out with the aid of a transmission line model \cite{pfeiffer2014bianisotropic}.
However, since this model neglects evanescent modes within the meta-atom, fine tuning via full-wave optimization is usually required to reach acceptable performance. 
To mitigate this hurdle, several circuit-based and field\textcolor{black}{-}based techniques have been proposed lately, allowing meta-atom synthesis without extensive use of full-wave solvers by incorporating near-field \emph{interlayer} coupling into the model \cite{xu2017technique,olk2019accurate,in proceedings URSI 2019}. 
While these allow accurate microscopic realization within the homogenization approximation, the macroscopic response may still deviate from the desired functionality due to mutual coupling between adjacent meta-atoms in the inhomogeneous MS, not properly accounted for in the design. 
More recently, we have addressed this \emph{intralayer} coupling issue as well, showing that our field-based model, rigorously considering interaction between \emph{all} individual scatterers in the \emph{macroperiod}, can be used to enhance the performance of the entire multilayer MS reliably without resorting to optimization in commercial full-wave simulators \cite{in proceedings ICEAA 2019}.

In this paper, we further demonstrate that proper extension of the detailed model can pave the path to rapid development of multifunctional MSs with high efficiency. Such MSs, allowing controlled response to several predefined excitations simultaneously with the same device, are attracting increasing interest in the last couple of years, enabling enhancements such as broad acceptance angle or large bandwidth, highly important for practical applications. 
While for reflect-mode MSs, using the aforementioned two-step macro/micro synthesis process may be feasible \cite{wang2020Independent}, the discussed practical design challenges would become more severe in transmit-mode cascaded formations. Building upon recent work on multilayer multielement metagratings \cite{rabinovich2019arbitrary}, we extend our previous work \cite{in proceedings ICEAA 2019} to include realistic dielectric substrates, enabling synthesis of practical fabrication-ready MSs. In particular, we show that by using a greedy strategy, we are able to iteratively optimize the individual scatterers for multiple design goals, corresponding to multiple possible angles of incidence for a given anomalous refraction MS. The ability to manipulate effectively a large number of degrees of freedom (scatterers) in conjunction with the high fidelity of the model forms a promising route for on-demand realization of advanced transmit-mode multifunctional devices.


\begin{figure}[htbp]
	\centering
	\includegraphics[width=14cm]{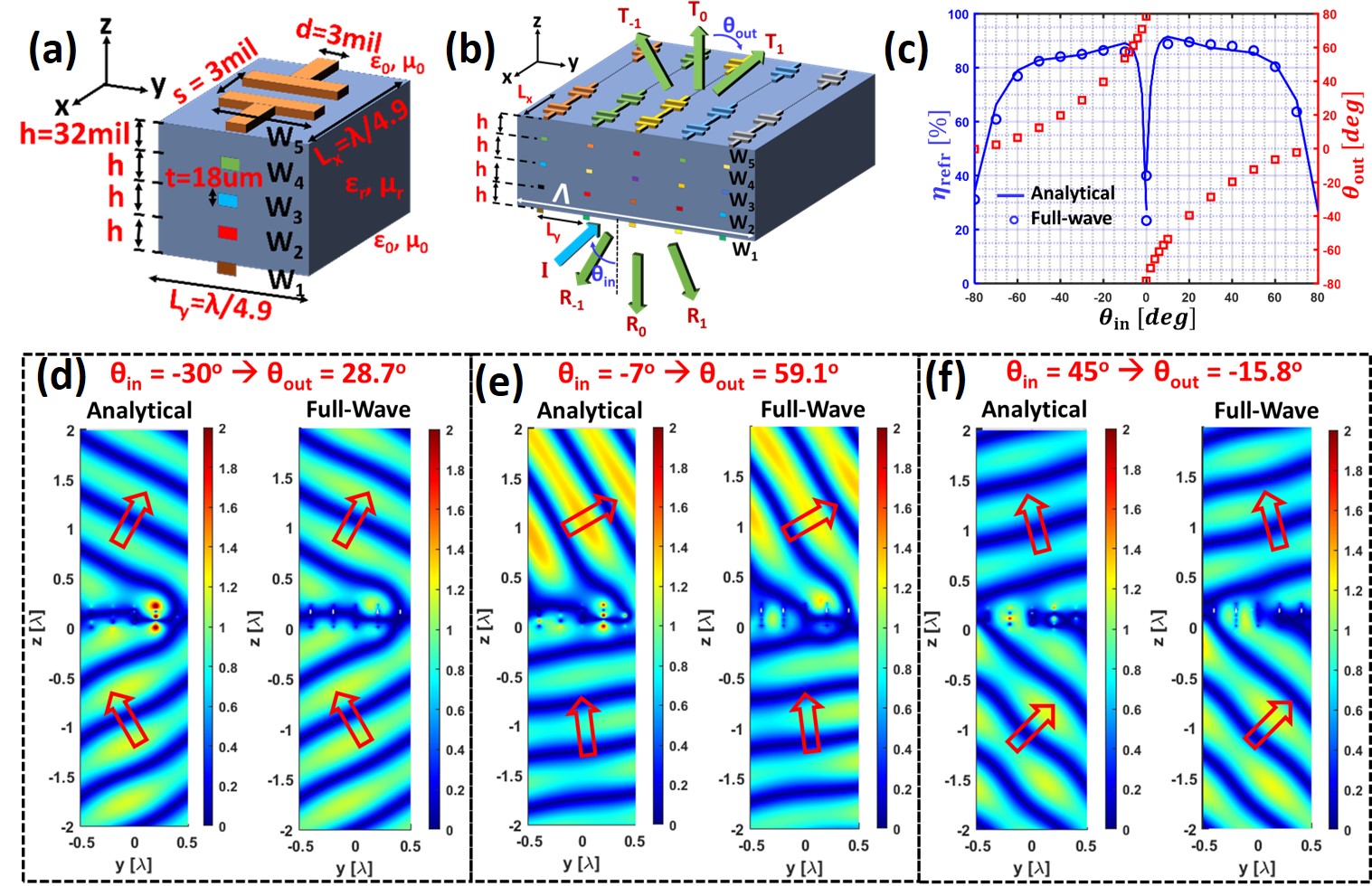}
	\caption{Schematic of the considered (a) meta-atom and (b) MS configurations ($\varepsilon_r=3$). 
	(c) Analytical (solid line) and full-wave (circle) predictions of refraction efficiencies (blue) for the multifunctional MS, along with the corresponding refraction angle $\theta_{\mathrm{out}}$ (red) as a function of $\theta_{\mathrm{in}}$. 
	Analytically calculated and full-wave simulated field distribution $|\Re\{E_x(y,z)\}|$ of the multifunctional MS for (d) $\theta_\mathrm{in}=-30^{o}$ (e) $\theta_\mathrm{in}=-7^{o}$, and (f) $\theta_\mathrm{in}=45^{o}$.
	} \label{Fig: cap}
\end{figure}

\section{Methodology, Results, and Discussion}
Our goal is to design a multifunctional MS implementing perfect anomalous refraction along the $\widehat{yz}$ plane for multiple excitation angles ($f=20$ GHz). To this end, we consider the MS configuration presented in Fig. \ref{Fig: cap}(b), excited by transverse electric fields ($E_z=E_y=H_x=0$). 
Each subwavelength $L_x\times L_y$ meta-atom [Fig. \ref{Fig: cap}(a)] consists of $N=5$ cascaded copper strips, loaded with printed capacitors of width $W_n$, with dielectric substrates ($\varepsilon_r=3$) of height $h$ in between. 
Since momentum transfer should only occur along $y$, we use $L_x=\lambda/4.9\ll\lambda$ as the period along $x$, effectively forming a 2D ($\partial/\partial x=0$) configuration \cite{rabinovich2019arbitrary}; a macro period $\Lambda=M\cdot L_y=(5/4.9)\lambda$ consisting of $M=5$ different meta-atoms is used along $y$ [Fig. \ref{Fig: cap}(b)], facilitating beam deflection. 

For such a given configuration, featuring $N\times M=25$ potentially different capacitors $W_{n,m}$ per period, and a given plane wave incidnet from $\theta_\mathrm{in}$, we can extend the analytical model presented in \cite{in proceedings ICEAA 2019} as per \cite{rabinovich2019arbitrary} to assess the fraction of power coupled to each of the scattered Floquet-Bloch (FB) modes. 
This is performed by calculating the induced currents in the various loaded strips, considering the effective load impedance associated with each of the printed capacitors, the incident field, and the secondary fields generated by the currents induced on the strips \cite{rabinovich2019arbitrary}. Subsequently, to optimize coupling to a specific mode, the first-order (anomalous refraction) in our case, we follow the greedy algorithm presented in \cite{in proceedings ICEAA 2019}. 
For each printed capacitor $(n,m)$, we sweep the capacitor width $W_{n,m}$ across the possible range, evaluating (using the analytical model) the scattering coefficients for all modes; the $W_{n,m}$ leading to the best coupling towards the first-order mode (including absorption) is retained, and we proceed to optimize the next capacitor in line. Repeating this procedure for all printed capacitors in the period yields one enhanced (potentially optimal) MS design. 
Since the choice of the \emph{initial} capacitor, from which this greedy optimization begins, affects the final layout and performance, we repeat this process 25 times, starting each time from a different capacitor within the period. 
Finally, 25 enhanced designs for anomalous refraction are \emph{semianalytically} obtained, from which the optimal one is chosen and subsequently verified in full-wave simulations.

For instance, we may apply this procedure to enhance anomalous refraction of a normally-incident wave ($\theta_\mathrm{in}=0$), starting from a simple phase-gradient MS (designed following \cite{in proceedings URSI 2019}) as an initial guess. 
For the MS of Fig. \ref{Fig: cap}(b), anomalous refraction would occur towards $\theta_\mathrm{out}=\arcsin\left(2\pi/\Lambda+\sin\theta_\mathrm{in}\right)=78.52^\circ$; prior to optimization, due to the large impedance mismatch \cite{epstein2016huygens}, the initial design yields refraction efficiency of $\eta_{\mathrm{refr}}=55.7 \%$. 
Applying the prescribed semianalytical scheme yields an enhanced design, predicted analytically to refract $\eta_{\mathrm{refr}} = 91.0\%$ of the incident power towards $\theta_\mathrm{out}=78.52^\circ$; full-wave simulations verify the enhancement, indicating $\eta_{\mathrm{refr}}= 86.1\%$\textcolor{black}{.} 
However, if now we consider excitation from a different angle, e.g. $\theta_\mathrm{in}=3^\circ$, the efficiency drops by about $5\%$; semianalytically optimizing the MS for \emph{this} incident angle now enhances the coupling to $\eta_{\mathrm{refr}}= 92.9 \%$ (predicted analytically), with the actual full-wave verified value being slightly smaller at $\eta_{\mathrm{refr}}= 90.7\%$.

Considering the efficacy of the presented methodology in adjusting the MS design as to enhance anomalous refraction for a given incident angle without requiring full-wave optimization, and in view of the large number of degrees of freedom available (25 capacitor widths $W_{n,m}$), we have decided to further harness it to synthesize multifunctional MSs.
Specifically, we aim at enhancing anomalous refraction efficiency for three different excitation scenarios \emph{simultaneously}, corresponding to angles of incidence $\theta^1_{\mathrm{in}}=-30^\circ$, $\theta^2_{\mathrm{in}}=-7^\circ$, and $\theta^3_{\mathrm{in}}=45^\circ$ ($\theta^1_{\mathrm{out}}=28.7^\circ$, $\theta^2_{\mathrm{out}}=59.1^\circ$, and $\theta^3_{\mathrm{out}}=-15.8^\circ$). 
To design a single MS that will be able to efficiently respond to all three excitations, we repeat the previously described greedy algorithm with one difference: when sweeping the capacitor widths $W_{m,n}$, we compute analytically the scattering coefficients three times (once for each of the excitation scenarios), and choose the width that maximizes all three (formally, maximizes $\min\{\eta^1_{\mathrm{refr}}, \eta^2_{\mathrm{refr}}, \eta^3_{\mathrm{refr}}\}$).

Executing the scheme as outlined leads to a fabrication-ready multifunctional MS design with predicted anomalous refraction efficiencies of $\eta^1_{\mathrm{refr}}=84.9 \%$,  $\eta^2_{\mathrm{refr}}=87.8 \%$, and $\eta^3_{\mathrm{refr}}=86.3 \%$ for the three incident angles, limited mostly by conductor loss. 
Defining these structures in ANSYS HFSS allows validation of the analytical calculations, with efficiencies of $\eta^1_{\mathrm{refr}}= 85.0 \%$,  $\eta^2_{\mathrm{refr}}= 83.1 \%$, and $\eta^3_{\mathrm{refr}}= 87.5 \%$ recorded in full-wave simulations [Fig. \ref{Fig: cap}(c)]. 
This agreement between theory and simulations is further emphasized in Fig. \ref{Fig: cap}(d), showing excellent correspondence between the analytically predicted field distributions and the ones obtained in HFSS. 
Remarkably, the anomalous refraction efficiency remains very high for a wide range of angles $\pm\theta_\mathrm{in}\in[4^\circ,60^\circ]$, indicating the immense potential of the proposed technique for designing devices with broad acceptance angle [Fig. \ref{Fig: cap}(c)]. 
Notably, this procedure was concluded in less than 7 minutes on a standard personal computer, directly generating a detailed PCB layout ready for manufacturing; experimental results will be reported in the symposium.

\section{Conclusion}
We have presented a semianalytical field-based method for designing multifunctional periodic MSs, avoiding time-consuming full-wave optimization. Properly incorporating near- and far-field mutual coupling, accounting for the detailed conductor geometry, dielectric substrate, and Ohmic losses, the model enables reliable synthesis under multiple design goals. As verified numerically, the fabrication-ready PCB MSs produced by this scheme exhibit high efficiencies for a broad range of excitation scenarios, forming a powerful tool for advanced MS engineering.



\end{document}